# A new inexact iterative hard thresholding algorithm for compressed sensing

Yuli Sun, Jinxu Tao *

*Department of Electronic Engineering and Information Science, University of Science and Technology of China, Hefei, Anhui 230027, People's Republic of China*

**ABSTRACT:** Compressed sensing (CS) demonstrates that a sparse, or compressible signal can be acquired using a low rate acquisition process below the Nyquist rate, which projects the signal onto a small set of vectors incoherent with the sparsity basis. In this paper, we propose a new framework for compressed sensing recovery problem using iterative approximation method via $\ell_0$ minimization. Instead of directly solving the unconstrained $\ell_0$ norm optimization problem, we use the linearization and proximal points techniques to approximate the penalty function at each iteration. The proposed algorithm is very simple, efficient, and proved to be convergent. Numerical simulation demonstrates our conclusions and indicates that the algorithm can improve the reconstruction quality.

## 1. INTRODUCTION

Compressed sensing is a technique to sample the sparse or compressible signals below the Nyquist rate, whilst still allowing perfect reconstruction of the signal [1]. CS has been attracting much attention over last few years due to its various potential applications. For example, a wide range of signal processing applications such as medical imaging [2], quantum-state tomography [3], radar systems [4] and communications [5] have benefited from progress made in CS.

Let $\mathbf{f} \in \mathbb{R}^N$ be the unknown signal, which usually obtained by vectorizing two-dimensional images or higher dimensional data into one-dimensional. Suppose that $\mathbf{f}$ has an expansion on the orthonormal basis $\mathbf{\Psi} = [\psi_1, \psi_2, \cdots, \psi_N] \in \mathbb{R}^{N\times N}$, which can be expressed as:

$$\mathbf{f} = \sum_{i=1}^{N} \psi_i x_i = \mathbf{\Psi x} \tag{1}$$

where $\mathbf{x} \in \mathbb{R}^N$ is the transform coefficient vector. $\mathbf{f}$ is said to be $K$-sparse (usually $K \ll N$) under $\mathbf{\Psi}$, if $\|\mathbf{x}\|_0$, the number of nonzeros in $\mathbf{x}$, is $K$.

More specifically, instead of sampling $\mathbf{f}$ using traditional sampling theory with $N$ samples, CS sample it with an $M \times N$ measurement matrix $\mathbf{\Phi} \in \mathbb{R}^{M\times N}$ and $M$ is smaller than $N$, yielding the measurement vector $\mathbf{y} \in \mathbb{R}^M$, which can be expressed as:

* Corresponding Author: Tel: +86-551-63601329; Email: tjingx@ustc.edu.cn.

$$\mathbf{y} = \mathbf{\Phi f} = \mathbf{\Phi \Psi x} = \mathbf{Ax} \tag{2}$$

where $\mathbf{A} = \mathbf{\Phi \Psi}$, may be called the compressed matrix [6].

Reconstruction of the signal $\mathbf{f}$ or, what is equivalent, the vector $\mathbf{x}$, is an underdetermined problem, which is usually formulated as following optimization problem:

$$\mathbf{x}^* = \arg\min_{\mathbf{x}} \|\mathbf{x}\|_p \quad s.t. \quad \mathbf{y} = \mathbf{Ax} \tag{3}$$

where $p$ is usually set to 1 or 0, $\|\mathbf{x}\|_1 = \sum_{i=1}^{N} |x_i|$ is the $\ell_1$ norm of $\mathbf{x}$, while $\|\mathbf{x}\|_0$ is the $\ell_0$ norm, counting the nonzero entries of $\mathbf{x}$. In this work we only consider the $\ell_0$ norm optimization problem.

Unfortunately, solving the above $\ell_0$ norm problem is known to be NP-hard in general [7]. The common approaches are solving the constrained optimization problem of the form:

$$(Q1): \ \mathbf{x}^* = \arg\min_{\mathbf{x}} \|\mathbf{y} - \mathbf{Ax}\|_2^2 \quad s.t. \quad \|\mathbf{x}\|_0 \le K \tag{4}$$

In words, they are looking for a vector $\mathbf{x}$, which have no more than $K$ nonzero coefficients, to minimizes the approximation error $\|\mathbf{y} - \mathbf{Ax}\|_2^2$. One type of these approaches is the greedy algorithms such as orthogonal matching pursuit (OMP) [8] and compressed sensing matching pursuit (CoSaMP) algorithm [9]. One advantage of the greedy approaches is that they can also be used to recover signals with more complex structures than sparsity, such as tree sparse signals [10]. Another type is the iterative hard thresholding (IHT) algorithms [11] such as normalized IHT [12] and accelerate IHT [13]. IHT is a simple algorithm and it has been proved that it can recover near-optimal solutions of the sparse signals under certain conditions [11]. Obviously, the reconstruction quality relies on the priori knowledge of sparsity $K$.

In this work, we consider the unconstrained optimization problem of (4):

$$(Q2): \ \mathbf{x}^* = \arg\min_{\mathbf{x}} \|\mathbf{x}\|_0 + \frac{\mu}{2} \|\mathbf{y} - \mathbf{Ax}\|_2^2 \tag{5}$$

where $\mu$ is a non-negative parameter. We proposed a new iterative algorithm to solve the above unconstrained $\ell_0$ norm optimization problem, which uses the linearization and proximal points techniques to approximate the second penalty function $\|\mathbf{y} - \mathbf{Ax}\|_2^2$ at each iteration. This inexact approximation method has shown its advantages in [14]-[15]. The proposed algorithm is very simple and we demonstrate its effectiveness by numerical examples.

The rest of the paper is structured as follows. In section 2, we propose the new $\ell_0$ norm reconstruction algorithm. In section 3, we analyze the properties and convergence of the proposed algorithm. Section 4 presents the numerical results. In the end, we provide our conclusion in section 5.

## 2. METHOD

The main difficulty in solving $(\mathcal{Q}2)$ comes from the $\ell_0$ norm minimization. Our approach is to approximate it with a simple function, which is easy to solve. Here, we rewrite the penalty function as:

$$\mathcal{L}^1(\mathbf{x},\mu)=\|\mathbf{x}\|_0+\frac{\mu}{2}\|\mathbf{y}-\mathbf{A}\mathbf{x}\|_2^2 \tag{6}$$

Firstly, suppose that we have $\mathbf{x}^k$ at the $k$-th iteration, then we expand the second penalty function $\|\mathbf{y}-\mathbf{A}\mathbf{x}\|_2^2$ at $\mathbf{x}^k$, that is:

$$\|\mathbf{y}-\mathbf{A}\mathbf{x}\|_2^2=\left\|\mathbf{A}\left(\mathbf{x}-\mathbf{x}^k\right)\right\|_2^2-2\left\langle\mathbf{A}^T\left(\mathbf{y}-\mathbf{A}\mathbf{x}^k\right),\left(\mathbf{x}-\mathbf{x}^k\right)\right\rangle+\left\|\mathbf{y}-\mathbf{A}\mathbf{x}^k\right\|_2^2 \tag{7}$$

The equation can be derived by:

$$\begin{aligned}\|\mathbf{y}-\mathbf{A}\mathbf{x}\|_2^2&=\|\mathbf{A}\mathbf{x}\|_2^2+\|\mathbf{y}\|_2^2-2\langle\mathbf{y},\mathbf{A}\mathbf{x}\rangle=\left\|\mathbf{A}\left(\mathbf{x}-\mathbf{x}^k\right)\right\|_2^2+\|\mathbf{y}\|_2^2-\left\|\mathbf{A}\mathbf{x}^k\right\|_2^2-2\left\langle\left(\mathbf{y}-\mathbf{A}\mathbf{x}^k\right),\mathbf{A}\mathbf{x}\right\rangle\\&=\left\|\mathbf{A}\left(\mathbf{x}-\mathbf{x}^k\right)\right\|_2^2+\left\|\mathbf{y}-\mathbf{A}\mathbf{x}^k\right\|_2^2-2\left\|\mathbf{A}\mathbf{x}^k\right\|_2^2+2\left\langle\mathbf{y},\mathbf{A}\mathbf{x}^k\right\rangle-2\left\langle\left(\mathbf{y}-\mathbf{A}\mathbf{x}^k\right),\mathbf{A}\mathbf{x}\right\rangle\\&=\left\|\mathbf{A}\left(\mathbf{x}-\mathbf{x}^k\right)\right\|_2^2-2\left\langle\mathbf{A}^T\left(\mathbf{y}-\mathbf{A}\mathbf{x}^k\right),\left(\mathbf{x}-\mathbf{x}^k\right)\right\rangle+\left\|\mathbf{y}-\mathbf{A}\mathbf{x}^k\right\|_2^2\end{aligned} \tag{8}$$

Besides, this expansion can be obtained by the multivariate Taylor formula.

We make an approximation of $\left\|\mathbf{A}\left(\mathbf{x}-\mathbf{x}^k\right)\right\|_2^2$ as:

$$\left\|\mathbf{A}\left(\mathbf{x}-\mathbf{x}^k\right)\right\|_2^2\approx\frac{1}{\tau^k}\left\|\mathbf{x}-\mathbf{x}^k\right\|_2^2 \tag{9}$$

where $\tau^k>0$ is a function of $x^k$. Then substitute (7) and (9) into (6), we can obtain

$$\mathcal{L}^2\left(\mathbf{x},\mu,\mathbf{x}^k,\tau^k\right)=\|\mathbf{x}\|_0+\frac{\mu}{2\tau^k}\left\|\mathbf{x}-\mathbf{x}^k\right\|_2^2-\mu\left\langle\mathbf{A}^T\left(\mathbf{y}-\mathbf{A}\mathbf{x}^k\right),\left(\mathbf{x}-\mathbf{x}^k\right)\right\rangle+\frac{\mu}{2}\left\|\mathbf{y}-\mathbf{A}\mathbf{x}^k\right\|_2^2 \tag{10}$$

To minimize $\mathcal{L}^2\left(\mathbf{x},\mu,\mathbf{x}^k,\tau^k\right)$ with respect to $\mathbf{x}$, we rewrite it as:

$$\begin{aligned}\min_{\mathbf{x}}\mathcal{L}^2\left(\mathbf{x},\mu,\mathbf{x}^k,\tau^k\right)&=\min_{\mathbf{x}}\left\{\|\mathbf{x}\|_0+\frac{\mu}{2\tau^k}\left\|\mathbf{x}-\left(\mathbf{x}^k+\tau^k\mathbf{A}^T\left(\mathbf{y}-\mathbf{A}\mathbf{x}^k\right)\right)\right\|_2^2\right\}+\frac{\mu}{2}\left\|\mathbf{y}-\mathbf{A}\mathbf{x}^k\right\|_2^2-\frac{\mu\tau^k}{2}\left\|\mathbf{A}^T\left(\mathbf{A}\mathbf{x}^k-\mathbf{y}\right)\right\|_2^2\\&=\min_{\mathbf{x}}\sum_{i=1}^N\left\{\|x_i\|_0+\frac{\mu}{2\tau^k}\left\|x_i-\left(x_i^k+\tau^k\left(\mathbf{A}_i\right)^T\left(\mathbf{y}-\mathbf{A}\mathbf{x}^k\right)\right)\right\|_2^2\right\}+\frac{\mu}{2}\left\|\mathbf{y}-\mathbf{A}\mathbf{x}^k\right\|_2^2-\frac{\mu\tau^k}{2}\left\|\mathbf{A}^T\left(\mathbf{A}\mathbf{x}^k-\mathbf{y}\right)\right\|_2^2\end{aligned} \tag{11}$$

where $\mathbf{A}_i\in\mathbb{R}^{M\times1}$ is the $i$-th column of $\mathbf{A}$. Instead of directly solving the difficult minimization of (6), we focus on its approximate minimization problem (11). And to solve this optimization problem, we further introduce a simple function:

$$\phi(a)=\|a\|_0+s(a-b)^2=\begin{cases}1+s(a-b)^2 & a\neq0\\ s\cdot b^2 & a=0\end{cases} \tag{12}$$

where $s>0$. We have

$$a^{*}=\arg\min_{a}\{\phi(a)\}=hard\left(b,\sqrt{1/s}\right)=\begin{cases}0 & |b|<\sqrt{1/s}\\ b & else\end{cases} \tag{13}$$

where $hard\left(b,\sqrt{1/s}\right)$ is the non-linear operator that sets all but the larger (in magnitude) than $\sqrt{1/s}$ elements of $b$ to zero.

Compare (12) with (11), we have the solution of (11):

$$for \quad i=1,2,\cdots,N$$

$$x_i^{k+1}=\arg\min_{x_i}\mathcal{L}^2\left(\mathbf{x},\mu,\mathbf{x}^k,\tau^k\right)=hard\left(x_i^k+\tau^k\left(\mathbf{A}_i\right)^T\left(\mathbf{y}-\mathbf{A}\mathbf{x}^k\right),\sqrt{2\tau^k/\mu}\right) \tag{14}$$

Then we obtain a new iterative algorithm for $(\mathcal{Q}2)$ and we call it as inexact iterative hard thresholding (IIHT) algorithm: initialize with $\mathbf{x}^0=0$ and use the iteration

$$x_i^{k+1}=hard\left(x_i^k+\tau^k\left(\mathbf{A}_i\right)^T\left(\mathbf{y}-\mathbf{A}\mathbf{x}^k\right),\sqrt{2\tau^k/\mu}\right) \tag{15}$$

IIHT in (15) biases the small entries towards zero. Here we consider a special and additional case where $\mathbf{A}$ is an orthogonal matrix (implying that $M=N$), which is not the main case of interest to us. In this case, there is no need to iterate, since (9) can be equal and the solution is a single step as $\mathbf{x}=hard\left(\mathbf{A}^T\mathbf{y},\sqrt{2/\mu}\right)$. As $\mathbf{y}$ represents a noisy measurement, IIHT removes the small entries of $\mathbf{A}^T\mathbf{y}$ but leaves large entries.

IIHT is very simple both in the iterative structure and the memory requirement. It only involves the application of the operators $\mathbf{A}$ and $\mathbf{A}^T$ once in each iteration as well as two vector additions. For large problems, the computational complexity can be reduced by using the structured operators such as fast Fourier transforms or wavelet transforms. The operator $hard(\bullet)$ involves a magnitude comparison of $\tilde{\mathbf{x}}^k=\mathbf{x}^k+\tau^k\mathbf{A}^T\left(\mathbf{y}-\mathbf{A}\mathbf{x}^k\right)$ with $\sqrt{2\tau^k/\mu}$. Apart from storage of $\mathbf{y}$, we only require the storage of the vector $\tilde{\mathbf{x}}^k$ and the nonzero elements of $\mathbf{x}^k$. So the storage requirement is small.

## 3. Convergence Analysis

In this section, we analyze the properties and convergence of the proposed IIHT.

**Theorem 1**. If $\mathbf{x}^{k+1}\neq\mathbf{x}^k$, then $\mathcal{L}^2\left(\mathbf{x}^{k+1},\mu,\mathbf{x}^k,\tau^k\right)\leq\mathcal{L}^2\left(\mathbf{x}^k,\mu,\mathbf{x}^k,\tau^k\right)$; only if $\left|\left(\mathbf{A}_i\right)^T\mathbf{y}\right|=\sqrt{2\tau^k/\mu}$ for all $i=1,2,\cdots,N$ and $\mathbf{x}^k=0$, then $\mathcal{L}^2\left(\mathbf{x}^{k+1},\mu,\mathbf{x}^k,\tau^k\right)=\mathcal{L}^2\left(\mathbf{x}^k,\mu,\mathbf{x}^k,\tau^k\right)$.

Proof:

Firstly, we study the function (12). If $a^e \neq a^*$, we can obtain: when $|b| \neq \sqrt{1/s}$, $\phi(a^*) < \phi(a^e)$; when $|b| = \sqrt{1/s}$, according to (13): $a^* = b$, then $\phi(a^* = b) = \phi(a^e = 0) = 1$.

Then, we compare (11) with (12), when $\left| x_i^k + \tau^k (\mathbf{A}_i)^T (\mathbf{y} - \mathbf{A}\mathbf{x}^k) \right| = \sqrt{2\tau^k / \mu}$ for all $i = 1, 2, \cdots, N$ and $\mathbf{x}^k = 0$, $\mathcal{L}^2(\mathbf{x}^{k+1}, \mu, \mathbf{x}^k, \tau^k) = \mathcal{L}^2(\mathbf{x}^k, \mu, \mathbf{x}^k, \tau^k)$.

**Theorem 2**. When $0 < \tau^k < \frac{1}{\|\mathbf{A}\|_2^2}$ and $\mathbf{x}^{k+1} \neq \mathbf{x}^k$, $\mathcal{L}^1(\mathbf{x}^{k+1}, \mu) < \mathcal{L}^1(\mathbf{x}^k, \mu)$.

Proof:

From (6) and (10), we can obtain

$$\mathcal{L}^2(\mathbf{x}, \mu, \mathbf{x}^k, \tau^k) = \mathcal{L}^1(\mathbf{x}, \mu) + \frac{\mu}{2}\left( \frac{1}{\tau^k} \left\| \mathbf{x} - \mathbf{x}^k \right\|_2^2 - \left\| \mathbf{A}(\mathbf{x} - \mathbf{x}^k) \right\|_2^2 \right) \tag{16}$$

If $0 < \tau^k < \frac{1}{\|\mathbf{A}\|_2^2}$ and $\mathbf{x}^{k+1} \neq \mathbf{x}^k$, $\mathcal{L}^1(\mathbf{x}, \mu) < \mathcal{L}^2(\mathbf{x}, \mu, \mathbf{x}^k, \tau^k)$.

Then we have $\mathcal{L}^1(\mathbf{x}^{k+1}, \mu) < \mathcal{L}^2(\mathbf{x}^{k+1}, \mu, \mathbf{x}^k, \tau^k)$ and $\mathcal{L}^1(\mathbf{x}^k, \mu) = \mathcal{L}^2(\mathbf{x}^k, \mu, \mathbf{x}^k, \tau^k)$. By using Theorem 1:

$$\mathcal{L}^1(\mathbf{x}^{k+1}, \mu) < \mathcal{L}^2(\mathbf{x}^{k+1}, \mu, \mathbf{x}^k, \tau^k) \leq \mathcal{L}^2(\mathbf{x}^k, \mu, \mathbf{x}^k, \tau^k) = \mathcal{L}^1(\mathbf{x}^k, \mu) \tag{17}$$

By using the above theorems, we have the convergence analysis as following:

**Theorem 3**. If $0 < \tau^k \leq \frac{1}{\|\mathbf{A}\|_2^2 + \delta}$ ($\delta$ is a small positive constant), $\lim_{k \to \infty} \left\| \mathbf{x}^{k+1} - \mathbf{x}^k \right\|_2 = 0$.

Proof:

Define $\varepsilon_k = \left\| \mathbf{x}^{k+1} - \mathbf{x}^k \right\|_2^2$, by using (16) and (17):

$$\begin{aligned} \mathcal{L}^1(\mathbf{x}^{k+1}, \mu) &\leq \mathcal{L}^1(\mathbf{x}^k, \mu) - \frac{\mu}{2}\left( \frac{1}{\tau^k} \left\| \mathbf{x}^{k+1} - \mathbf{x}^k \right\|_2^2 - \left\| \mathbf{A}(\mathbf{x}^{k+1} - \mathbf{x}^k) \right\|_2^2 \right) \leq \mathcal{L}^1(\mathbf{x}^k, \mu) + \frac{\mu\varepsilon_k}{2}\left( \|\mathbf{A}\|_2^2 - \frac{1}{\tau^k} \right) \\ &\leq \mathcal{L}^1(\mathbf{x}^k, \mu) - \frac{\mu\delta\varepsilon_k}{2} \leq \mathcal{L}^1(\mathbf{x}^0, \mu) - \frac{\mu\delta}{2}\sum_{i=0}^{k} \varepsilon_i = \frac{\mu}{2}\|\mathbf{y}\|_2^2 - \frac{\mu\delta}{2}\sum_{i=0}^{k} \varepsilon_i \end{aligned} \tag{18}$$

Because $\mathcal{L}^1(\mathbf{x}, \mu) \geq 0$, then we have

$$\varphi_k = \sum_{i=0}^{k} \varepsilon_i \leq \frac{\|\mathbf{y}\|_2^2}{\delta} \tag{19}$$

As $\varphi_k$ is a monotonically increasing sequence and upper bounded, then it is a convergent sequence and $\lim_{k\to\infty}\left\|\mathbf{x}^{k+1}-\mathbf{x}^{k}\right\|_2 = 0$.

## 4. Simulation

In this section, simulation is performed to demonstrate the proposed conclusions and evaluate the performance of the IIHT algorithm.

We apply five methods to reconstruct the images: (1) direct measurement $\mathbf{x}=\mathbf{A}^T\mathbf{y}$; (2) the proposed IIHT using the $\ell_0$ norm minimization; (3) compressed sensing matching pursuit (CoSaMP); (4) iterative hard thresholding (IHT) algorithm; (5) iterative soft thresholding (IST) algorithm [16] using the $\ell_1$ norm minimization using the iteration as :

$$\mathbf{x}^{k+1} = \mathbb{S}_{w^k}\left(\mathbf{x}^k + \tau\mathbf{A}^T\left(\mathbf{y}-\mathbf{A}\mathbf{x}^k\right)\right);\ \left(\mathbb{S}_{w^k}\left(\mathbf{x}\right)\right)_i = \begin{cases} x_i - w^k\operatorname{sign}\left(x_i\right) & \left|x_i\right| \geq w^k \\ 0 & \left|x_i\right| < w^k \end{cases} \tag{20}$$

And we easily set $w^k = \frac{\left\|\mathbf{x}^k + \tau\mathbf{A}^T\left(\mathbf{y}-\mathbf{A}\mathbf{x}^k\right)\right\|_1 - \left\|\hat{\mathbf{x}}\right\|_1}{N}$ to make $\left\|\mathbf{x}^{k+1}\right\|_1$ close to the $\ell_1$ norm of original signal $\hat{\mathbf{x}}$. All experiments are performed in MATLAB v7.8 (2009a) running on a Lenovo laptop with Intel(R) Core(TM) 2 Duo CPU P7350 (2.0GHZ), 2.0GB memory and Windows 7 operating system.

In the noisy model, $\mathbf{e}$ is the additive Gaussian noise generated by $\mathbf{e}=\sigma\cdot\overline{|y|}\cdot\operatorname{randn}\left(M,1\right)$, where $\overline{|y|} = \frac{\sum_{i=1}^{M}|y_i|}{M}$ is the average value of $\mathbf{y}$ magnitude. The process is started at the directly reconstructed $\mathbf{x}^0=\mathbf{A}^T\mathbf{y}$, and terminated when it reaches the max interactions number $N_{iter}$ or the $\ell_2$ distance between two successive reconstructions is small enough, that is:

$$\frac{\left\|\mathbf{x}^{k+1}-\mathbf{x}^{k}\right\|_2}{\left\|\mathbf{x}^{k}\right\|_2} \leq \text{tol}$$

which means that there is no longer any appreciate changes in the iteration and the algorithm runs into convergence. The quality of reconstructed vector is measured by the peak signal to noise ratio (PSNR) and relative error (Rel.Err) to the original signal $\hat{\mathbf{x}}$:

$$\text{Rel.Err} = \frac{\left\|\mathbf{x}-\hat{\mathbf{x}}\right\|_2}{\left\|\hat{\mathbf{x}}\right\|_2}\times 100\%$$

In the test, we use the Gaussian matrix $\mathbf{A}$ whose elements are generated from i.i.d. normal distribution $\mathcal{N}\left(0,1\right)$, and use $\text{SR}={M}/{N}$ to denote the sampling ratio. Since it is hard to know the $\left\|\mathbf{A}\right\|_2^2$ in advance for a large random $\mathbf{A}$, we set $\tau^k = \min\left\{\frac{1}{\left\|\mathbf{y}-\mathbf{A}\mathbf{x}^k\right\|_2^2}, \left\|\mathbf{y}-\mathbf{A}\mathbf{x}^k\right\|_2^2\right\}$ for IIHT. This

makes the step size $\tau^k$ to be small in the beginning and end, large in the middle. Meanwhile, the convergence speed is fast in the middle and slow in the two ends.

In the first study, we use random vectors $\hat{\mathbf{x}}$ with length $N = 2^{14}$ and sparsity levels $\text{SL}= K/N = 0.05$ as the original signals to be recovered. We initialize the parameters as $\text{SR=0.35}$, $N_{iter} = 100$, and $\text{tol} = 10^{-5}$.

We compare the reconstruction quality of different methods under different noise levels, as shown in Table 1. We set $\mu$ to be 350 and 170, corresponding to the $\sigma = 10\%$ and $\sigma = 20\%$, respectively. From Table 1, one can find that the proposed IIHT can make a great improvement when comparing with the direct measurement, and sometimes seems better than other three algorithms. This shows the effectiveness of the IIHT.

**Table 1**. Performance comparison of different algorithms.

| Parameters | | Measurement | IST | CoSaMP | IHT | IIHT |
|---|---|---|---|---|---|---|
| $\sigma = 10\%$ | Rel.Err | 165.69% | 12.52% | 6.61% | 3.94% | 3.81% |
| | PSNR(dB) | 25.25 | 47.68 | 53.24 | 57.72 | 58.16 |
| $\sigma = 20\%$ | Rel.Err | 176.89% | 18.49% | 14.31% | 9.27% | 8.21% |
| | PSNR(dB) | 25.04 | 44.65 | 46.88 | 50.65 | 51.71 |

In the second study, we use a sparse $128 \times 128$ ellipse phantom as the original $\hat{\mathbf{x}}$, as shown in Fig. 1, has a sparsity of 1282 pixels and $\text{SL}= K/N \approx 0.08$. This phantom is similar as in [17], which can be used in boundary-enhanced X-ray phase-contrast tomography. We initialize the parameters as $\mu = 2^8$, $\text{SR=0.35}$, $\sigma\text{=8\%}$, $N_{iter} = 800$, and $\text{tol} = 10^{-5}$. After trying different choices of parameters for IHT and IST, we select $\tau$=0.5 for IHT, $\tau$=0.3 for IST, which make they convergent and provide better performance.

Figure 1 shows the reconstructions of these methods, and in order to highlight the differences, Figure 2 shows the absolute differences relative to the phantom image. Comparing Fig. 1f and Fig. 2d with other reconstructed images, one can find that the proposed IIHT algorithm can achieve a high accuracy and competitive reconstruction. This once again shows the effectiveness of IIHT algorithm.

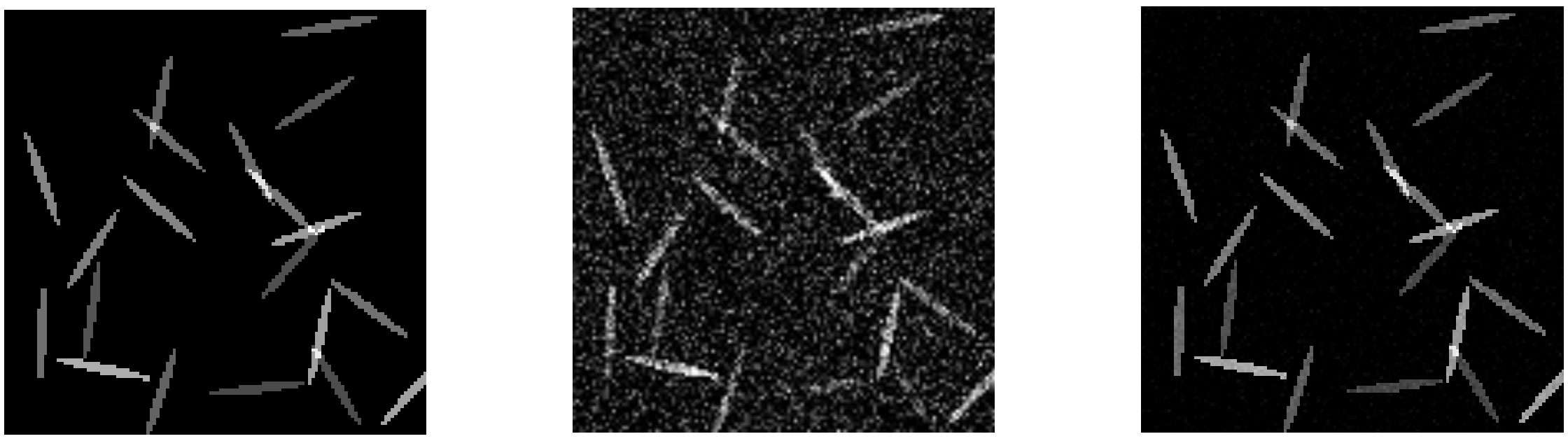

(a) (b) Rel.Err: 169.98%; PSNR: 17.96 (c) Rel.Err: 9.64%; PSNR: 42.89

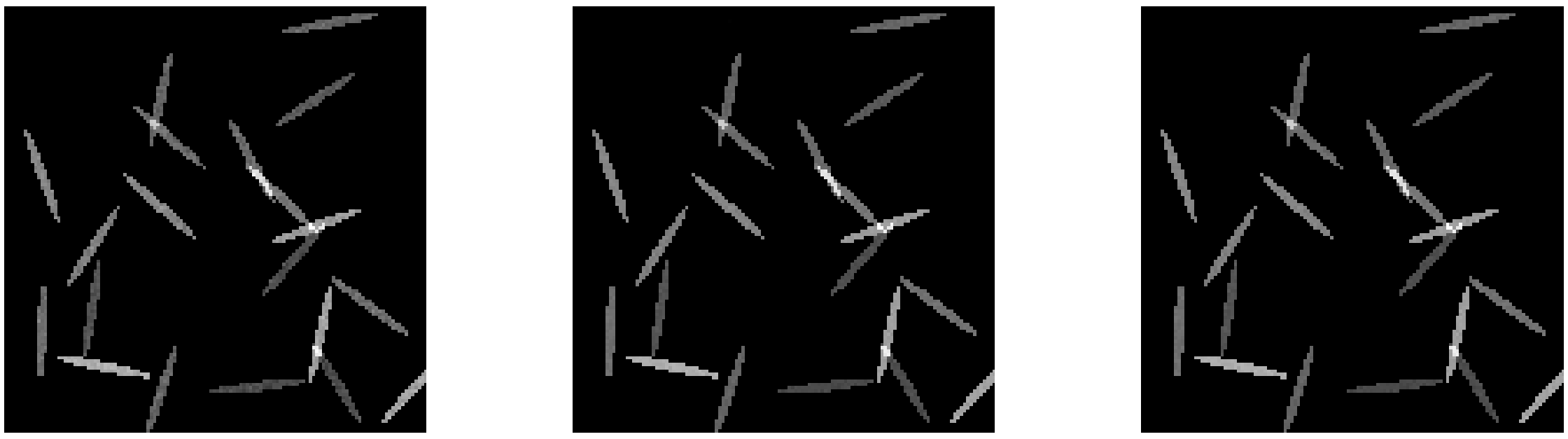

(d) Rel.Err: 6.98%; PSNR: 45.68 (e) Rel.Err: 3.48%; PSNR: 51.72 (f) Rel.Err: 3.47%; PSNR: 51.74

**Figure 1.** Reconstructed images of the ellipse phantom. (a) is the original phantom. From (b) to (f), corresponding to the reconstructions using direct measurement, IST, CoSaMP, IHT and IIHT, respectively. The display window is [0, 1.1].

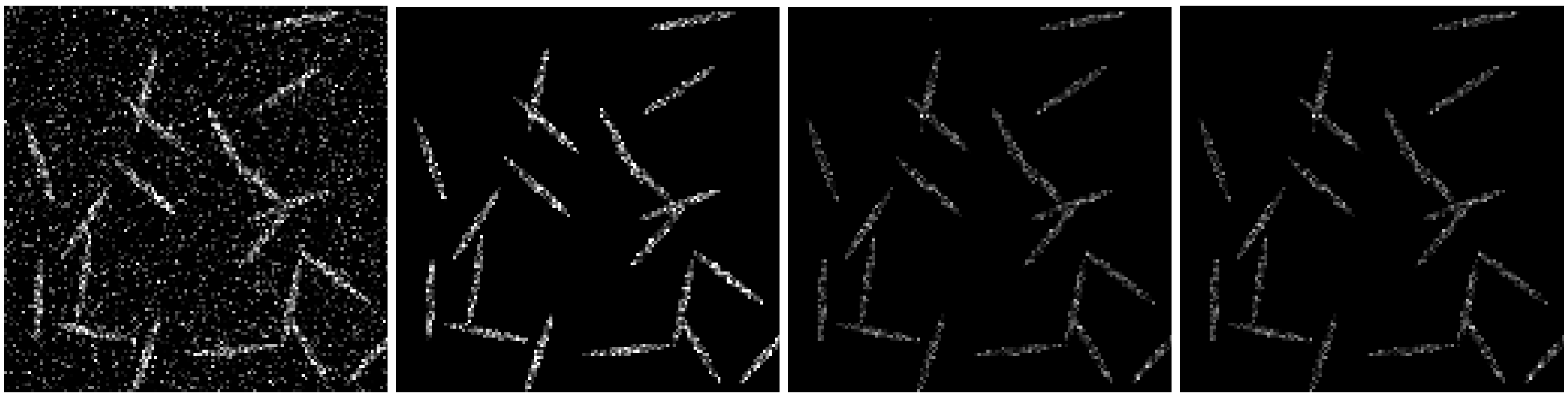

(a) (b) (c) (d)

**Figure 2.** Absolute differences relative to the phantom image. From (a) to (d), corresponding to the absolute differences relative to the phantom image of the reconstructions using IST, CoSaMP, IHT and IIHT, respectively. The display window is [0, 0.07].

## 5. Conclusion

In this paper, we propose a new iterative algorithm for compressed sensing recovery based on $\ell_0$ minimization. Since directly solving the unconstrained $\ell_0$ norm optimization problem is known to be hard, we use a new penalty function, which is easy to solve, to approximate it at each iteration. The proposed algorithm is very simple, efficient, and proved to be convergent. The simulation shows the effectiveness of this new algorithm. However, the optimal step size of the algorithm is still unsolved for us, as a better choice of step size can accelerate the algorithm, which is also our next work. In the near future, we will evaluate the algorithm with actual applications such as the few views reconstruction in computer tomography (CT), and we believe that the proposed algorithm is expected to have potential practical merits.

## References

[1] E.J. Candes, J. Romberg and T. Tao, "Robust Uncertainty Principles: Exact Signal Reconstruction from Highly Incomplete Frequency Information," *IEEE Trans. Inf. Theory*, vol.52, no. 2, Feb. 2006, pp. 489-509.

[2] M. Chang et al., "A Few-view Reweighted Sparsity Hunting (FRESH) Method for CT Image Reconstruction," *J. X-Ray Sci. Technol.*, vol. 21, no. 2, 2013, pp. 161-176.

[3] D. Gross et al., "Quantum State Tomography via Compressed Sensing," *Phys. Rev. Lett.*, vol. 105, no. 15, Oct. 2010, pp. 150401.

[4] V.M. Patel et al., "Compressed Synthetic Aperture Radar," *IEEE J-STSP.*, vol. 4, no. 2, Apr. 2010, pp. 244-254.

[5] C.R. Berger et al., "Sparse Channel Estimation for Multicarrier Underwater Acoustic Communication: From Subspace Methods to Compressed Sensing," *IEEE Trans. Signal Process.*, vol. 58, no. 3, Mar. 2010, pp. 1708-1721.

[6] A.M. Rateb and S.K. Syed-Yusof, "Performance Analysis of Compressed Sensing Given Insufficient Random Measurements," *ETRI J.*, vol. 35, no. 2, Apr. 2013, pp. 200-206.

[7] B.K. Natarajan, "Sparse Approximate Solutions to Linear Systems," *SIAM J. Comput.*, vol. 24, no. 2, Apr. 1995, pp. 227–234.

[8] D. Needell and R. Vershynin, "Uniform Uncertainty Principle and Signal Recovery via Regularized Orthogonal Matching Pursuit," *Found. Comput. Math.*, vol. 9, no. 3, June 2009, pp. 317–334.

[9] D. Needell and J. Tropp, "COSAMP: Iterative Signal Recovery from Incomplete and Inaccurate Samples," *Appl. Comput. Harmon. A.*, vol. 26, no. 3, May 2008, pp. 301–321.

[10] R.G. Baraniuk et al., "Model-based Compressive Sensing," *IEEE Trans. Inf. Theory*, vol. 56, no. 4, Apr. 2010, pp. 1982-2001.

[11] T. Blumensath and M.E. Davies, "Iterative Hard Thresholding for Compressed Sensing," *Appl. Comput. Harmon. A.*, vol. 27, no. 3, Nov. 2009, pp. 265-274.

[12] T. Blumensath and M.E. Davies, "Normalized Iterative Hard Thresholding: Guaranteed Stability and Performance," *IEEE J-STSP.*, vol. 4, no. 2, Apr. 2010, pp. 298-309.

[13] T. Blumensath, "Accelerated Iterative Hard Thresholding," Signal Process., vol. 92, no. 3, Mar. 2012, pp. 752-756.

[14] B. He et al., "A New Inexact Alternating Directions Method for Monotone Variational Inequalities," *Math. Program.*, vol. 92, no. 1, Mar. 2002, pp. 103-118.

[15] Y.H. Xiao and H.N. Song, "An Inexact Alternating Directions Algorithm for Constrained Total Variation Regularized Compressive Sensing Problems," *J. Math. Imaging Vis.*, vol. 44, no. 2, Oct. 2012, pp. 114-127.

[16] I. Daubechies, M. Defrise and C.D. Mol, "An Iterative Thresholding Algorithm for Linear Inverse Problems with A Sparsity Constraint," *Commun. Pur. Appl. Math.*, vol. 57, no. 11, Nov. 2004, pp. 1413-1457.

[17] E.Y. Sidky, M.A. Anastasio and X. Pan, "Image Reconstruction Exploiting Object Sparsity in Boundary-enhanced X-ray Phase-contrast Tomography," *Opt. Express*, vol. 18, no. 10, May 2010, pp. 10404-10422.